# Polar regions activity and the prediction of the height of the solar cycle 25

S. Koutchmy[1], B. Filippov[2], E. Tavabi[3], J-C. Noens[4], O. Wurmser[5]

[1] Institut d'Astrophysique de Paris, CNRS and Sorbonne University, UMR 7095- 98 Bis Boulevard Arago, 75014 Paris, France e-mail: koutchmy@iap.fr
[2] Pushkov Institute of Terrestrial Magnetism, Ionosphere and Radio Wave Propagation, Russian Academy of Sciences, Troitsk, Moscow, 108840, Russia
[3] Physics Department, Payame Noor University (PNU), 19395-3697, Tehran, I. R. of Iran
[4] Obs. Pic du Midi, Association des O.A., France
[5] Association des O.A., PdM, France

**ABSTRACT**

The forthcoming solar cycle (SC) 25 was beleived to be rather low when using the sunspot number (SN) as a measurement of the level of activity. The most popular prediction was made by the panel of NASA in 2019, including works based on extrapolations of dynamo-type models. We however discovered that using different observations to measure the level of polar regions activity several years before the start of SC25 and also after the start of the SC25 in 2020, the height of the SC25 could be high. The polar regions activity we considered seems related to the polar coronal holes (CH) activity and it is found significantly higher before the SC25 that it was before the SC24 and accordingly, we suggest that the SN cycle could indeed be much higher than during the SC24 that was a low SN height cycle.

## 1 Introduction

The solar activity through the sunspot number (SN) modulates geomagnetism, producing CMEs, flares, SEPs and associated disturbances. The prediction of the height of the forthcoming SC 25 expressed by the so-called SN is since 2019 the subject of many studies. These studies are mainly based on statistical and/or mathematical and/or heuristic methods, taking parameters from the analysis of past cycles; they leaded to a predicted SC 25 similar to the low height preceding SC 24, sometimes significantly lower, indeed. Some predictions uses solar activity parameters justified by the solid belief that the solar activity is fully governed by a dynamo mechanism occurring inside the Sun (see e.g. (Kitiashvili, 2020) and her reproduced from NASA graphics of SC 23 to 25 in Fig. 1). The popular dynamo model is the Babcock-Leighton (B-L) model describing the transformation of a general poloidal field of the rotating Sun into a toroidal field through the differential rotation visualized near the surface; it rather well reproduces the behavior of sunspots of different sign for each hemisphere during a SC. The regeneration of the new poloidal field (after 11 years) is another aspect of the model that is left not well understood despite many attempts to guess what is going on in polar regions during the cycle. The reversal of the dominant polarities in polar regions is since 1970 observed to occur during the years of maximum SN. However several puzzling features like the M-regions, the active longitudes, the occurrence of long-live big single sunspots, the cyclonic and the widely distorted behavior of extended interacting active regions, the occurrence of Coronal Holes not predictable by the dynamo effect, the polar regions cycles and/or field-reversals, and finally the large dispersion of heights of SN cycles etc. are however the subject of hot debates. More important for practical obvious reasons is the prediction of the solar cycles in advance (Nandy, 2021)

## 2 The polar regions activity during the SCs

It has been naively suggested (following the Ohl's law) that the Polar Regions activity, with the occurrence of recurrent geo-activity in the years around the solar minimum of SC *n* is rather correlated with the height of the *n +1* SC. Accordingly the height of the following SC could be predicted; additionally, there is now a growing consensus on the key role of polar magnetic fields as seeds for the SC (Makarov et al ., 1987). Finally, there is a suggestion that the chromospheric shell of cool lines changes of height or of thickness in polar regions during the SC see (Filippov et al., 2007) making the chromosphere oblate at SN minimum and prolate during the years of S$_N$ maximum such that it would reflect the influence of a global magnetic field during the SC.

### 2.1 Polar faculae, CHs and polar magnetic fields variations

We looked at the activity of Polar Regions using proxies like the density and brightness of polar faculae (small-scale W-L brightenings of several hours duration). Their apparent intensities or better, their radiative fluxes and their number were visually evaluated from HMI of SDO mission W-L full disk filtergrams after compensating the center-limb variations. This was done during the years of SN minimum of SC23 and SC24. Some research was also done using similar filtergrams from the MDI of the SOHO mission in years of the SC22. Unfortunately, we discovered that the analysis could be biased by the variable during the years and decades quality of filtergrams mainly due to the slightly changing quality of the focusing near the limbs. We can just qualitatively report the evaluation of the apparent number of polar faculae that shows a definite lower number before the SC24 and a seemingly more contrasted fluxes with a larger number before the SC24 and also during the first years of SC25. This is especially apparent in the polar coronal holes (not seen in W-L) when the solar B-angle is favorable (roughly in March- April for the S-pole and September-October for the N-pole). Because we believe the polar faculae activity reflects the behavior of the polar region magnetic field through the so-called reconnection phenomena, it is interesting to compare this behavior using the unique measurements of net (mean) polar large scale magnetic fields done at the Stanford WSO for several SC, see Fig. 2 and http://wso.stanford.edu/#MeanField. Nothing really significant seems to emerge from the recorded behavior of the strength of the mean fields but a general decreasing tendency along the cycles since 1975 when such measurements were started. It could be necessary to examine the synoptic maps in details to go further. The impression is that during the last SC these measurements could possibly suffer from an instrumental drift producing a weaker calibrated signal and it would be important to have another series from another observatory to check the conclusions. It is also possible that the signals during the first cycles were better "extracted"?

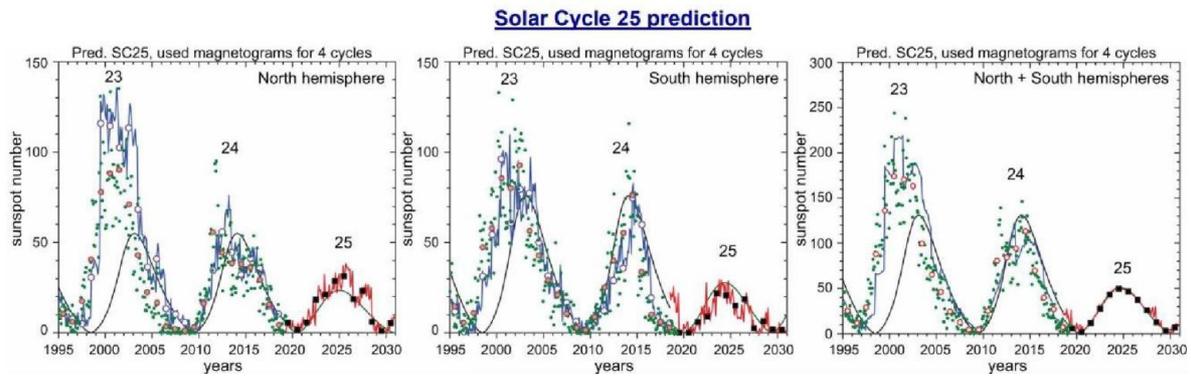

**Fig. 1.** Past recent SC of SN and extrapolated SC25 based on solar dynamo model by I. Kitiashvili, from a NASA document; see also (Kitiashvili, 2020).

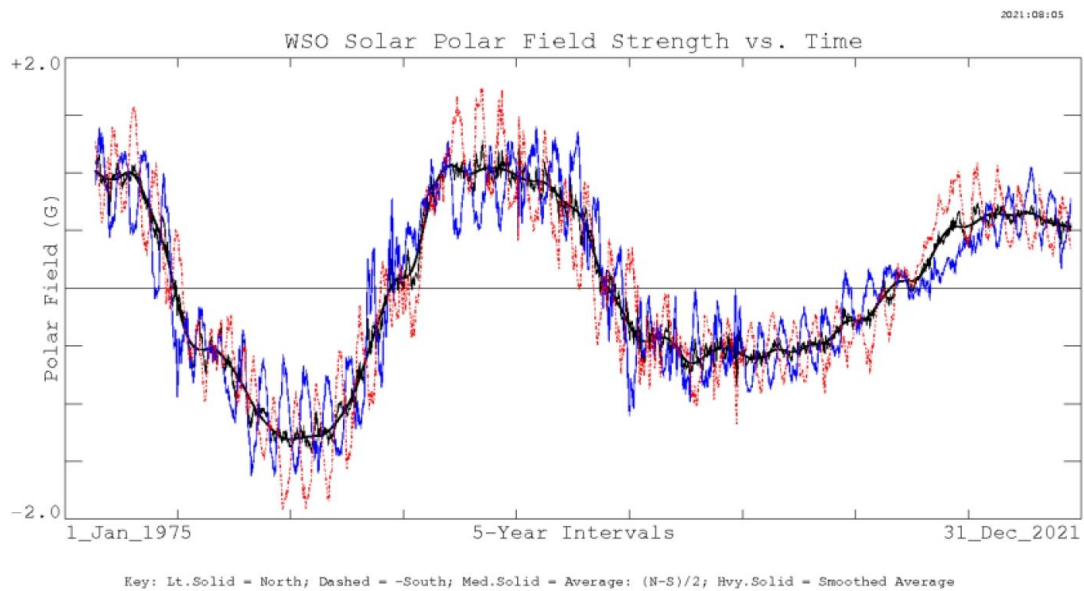

**Fig. 2.** Strength of the net mean Polar magnetic fields measured during the last SCs from the Stanford mapping of large scale magnetic field fluxes. Note the yearly periodicity due to the influence of the B-angle variations for each hemisphere.

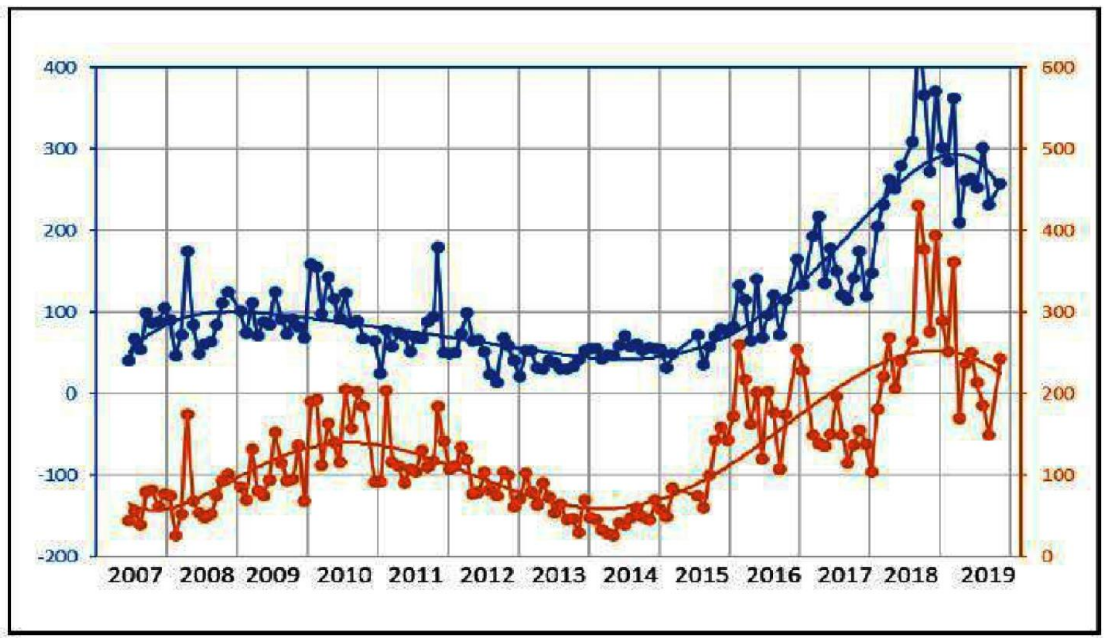

**Fig. 3.** Pic du Midi Observations of ejections in *Ha* during the last SC24. In blue for the N-pole and in brown for the S-pole. Values in ordinate are the monthly numbers.

**2.2 Ha ejection activity recorded at the Pic du Midi Observatory during SC24**

During the last SC24 the Pic du Midi Observatory coronagraph working with an Ha filter was used all along the years to observe the "cool" corona with good resolutions permitting as well to measure the number of small ejections events above the occulted by the coronagraph limb the spic- ular fringe of typically 7″ to 10" thickness. These events are usually called ejections or spikes, or macro-spicules (see ( Noens and Wurmser, 2000) for a description of the method used and for more detailed results). In Fig. 3 we show the most significant results, reported for each solar hemisphere. There is a definite change in amplitudes of the behavior of the number of polar events observed along the years of the SC24 when the period 2009-2011 (years before the SC24) is compared with the period 2018-2020 (years before SC25), typically by a factor 2 in favor of the years just before the SC25. Note that the team at the Pic du Midi Observatory checked their results several times for any instrumental bias they could have without finding a reasonable explanation. The effect looks of solar origin.

**2.3 Polar regions HeII 304 A activity from AIA of SDO mission images**

Extensions in polar regions we additionally studied are believed to be due to ejection events at transition region temperatures in nearly radial directions. They were called macro-spicules in the time of the SkyLab observations made in CIV line emissions, in contrast to spicules seen lower everywhere around the disk at lower temperatures, including regions outside the coronal hole regions. High quality AIA

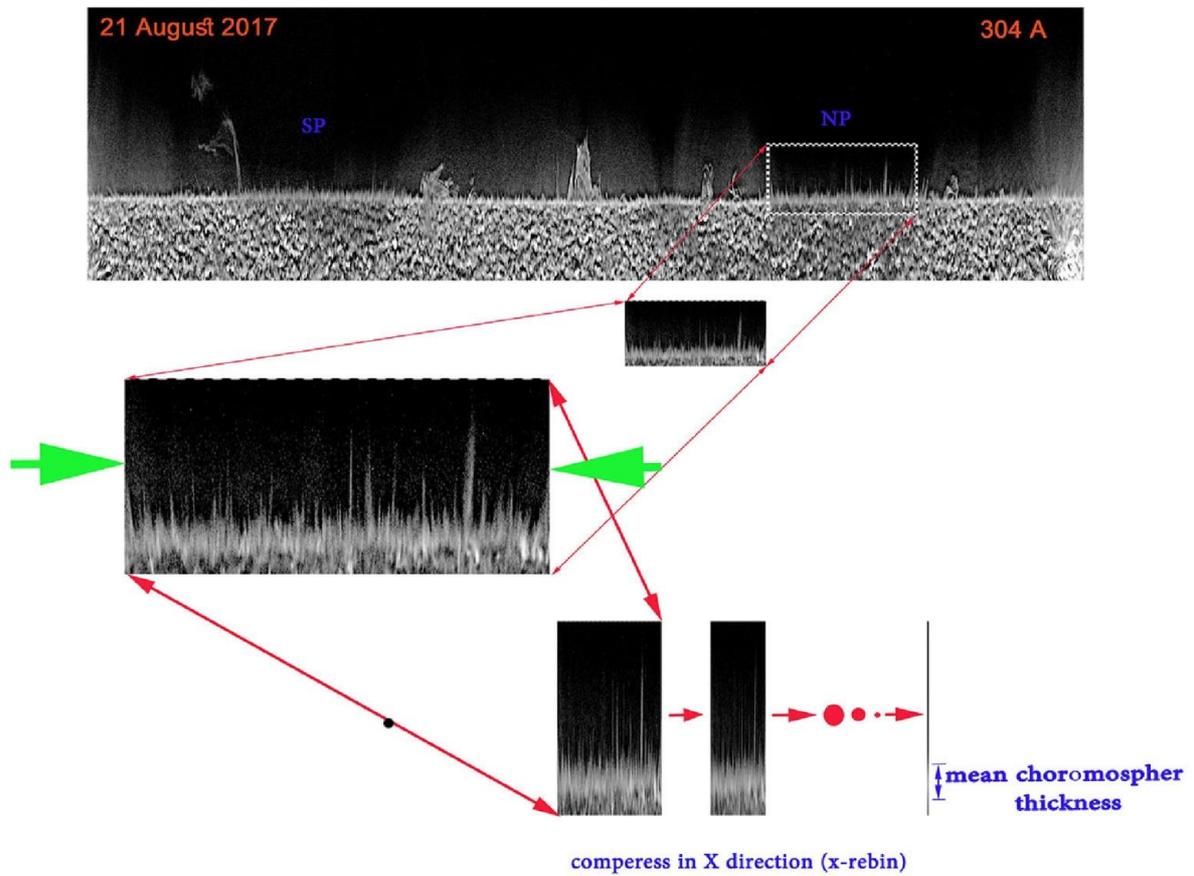

**Fig. 4.** Partial frame reconstructed image of a typical polar region to demonstrate the method used for measuring the "abnormal" thickness of the 304 A emissions (due to the HeII resonance line formed around 50 000K) in this polar region visualized in polar coordinates after averaging (by summing) the 304 A AIA images of SDO for 10 min. Such extensions were measured during more than one solar cycle in both the South and North regions.

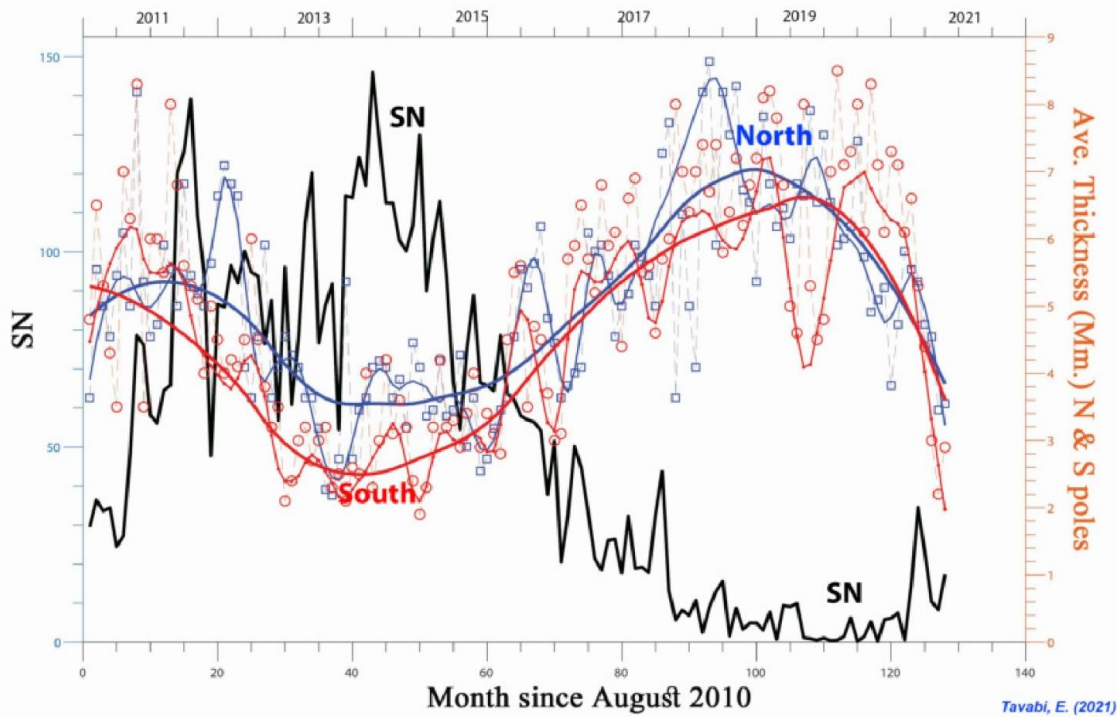

**Fig. 5.** Polar activity variations evaluated in the HeII 304 A emissions above the limb on summed images similar to image of Fig. 4, using the measurement of the "thickness" (FWHM) of the shell in radial directions.

304 A filtergrams of the HeII resonance line from the SDO NASA mission are now available and they were used after summing original frames for 10 min in order to improve the signal/noise ratio. We also worked in polar coordinates in order to more easily and more precisely measure the effective thickness (FWHM) of the polar regions of 304 A emissions, including the so-called "abnormal" thickness of the fringe above CHs, see Fig. 4 for an illustration of the different steps used to deduce these FWHM.

In Fig. 5 we show the results of measuring the FWHM of the 304 A shell for the last SC24. Again in both hemisphere we recorded a more important activity reflected by the FWHM thickness of the 304 A shell before the SC25 compared to the level recorded before the SC24.

## 3 Conclusion

We looked at the activity of Polar Regions using proxies: i/ density of polar faculae from visually evaluated HMI of SDO mission W-L filtergrams and the Stanford WSO large scale magnetograms evaluating polar regions magnetic net fluxes; ii/ numbers of cool ejection events from a 15 year survey of the Pic du Midi CLIMSO Ha observations ( Noens and Wurmser, 2000); iii/ averaged extensions of the 304 A shell in Polar Regions related to the polar CHs macro-spicules activity. Time variations of these 3 parameters qualitatively point to a SC25 that could reach high levels, of order of 2 times the height of the SC24, in contrast with the moderate height predicted by the SC25 Prediction Panel of NASA and NOAA (Chair: Doug Biesacker). The reason of this discrepancy is not clear. We better wait the occurrence of the (double) maximum of SC25 in 2025-26 to go further with the interpretation. Another interesting parameter seemingly related to this topic is the definite observation of the chromospheric prolateness (ovalisation) in the years of the minimum of 2018-2020 that was discovered in the years 1998-2020 (before SC 23) and that was not well measured in 2010-11 (before SC24). In rather "cool" spectral lines like Ha or the H and K lines of CaII, the smoothed upper edge of the solar chromosphere is prolate in the North-South direction at the epoch of minimum solar activity (Filippov and Koutchmy, 2000).